\documentclass[aps,preprint]{revtex4}%
\usepackage{amsfonts}
\usepackage{amsmath}
\usepackage{amssymb}
\usepackage{graphicx}%
\begin{document}
\title{\textbf{Entropy Conservation in the Transition of Schwarzschild-de Sitter
space to de Sitter space through tunneling}}
\author{ZHANG Bao-Cheng 
$^{1,2}$}
\email{zhangbc@wipm.ac.cn}
\author{CAI Qing-Yu 
$^{1}$}
\author{ZHAN Ming-Sheng 
$^{1,2}$}
\affiliation{$^{1}$State Key Laboratory of Magnetic Resonances and Atomic and Molecular
Physics, Wuhan Institute of Physics and Mathematics, The Chinese Academy of
Sciences, Wuhan 430071, China}
\affiliation{$^{2}$Center for Cold Atom Physics, Chinese Academy of Sciences, Wuhan 430071, China}

\begin{abstract}
We revisit the Parikh-Wilczek tunneling through the de Sitter horizon and
obtain the tunneling rate in Schwarzschild-de Sitter space, which is
non-thermal and closely related to the change of entropy. We discuss the
thermodynamics of Schwarzschild-de Sitter space and show existence of
correlation which ensured the conservation of the total entropy in the
transition process of Schwarzschild-de Sitter space to de Sitter space. The
correlation and the conserved entropy enlighten a way to explain the entropy
in empty de Sitter space.

PACS classification codes: 04.70.Dy, 04.60.-m, 97.60.Lf


\end{abstract}
\maketitle

\section{\textbf{Introduction}}

In 1974, Hawking discovered \cite{swh75} that when considering quantum effect,
a black hole can emit thermal radiation with a temperature $T=\frac{\kappa
}{2\pi}$, where $\kappa$ is the surface gravity of the black hole. The
physical reason of radiation was explained \cite{hh76} as coming from vacuum
fluctuations tunneling through the horizon of the black hole. But some
original derivations \cite{swh75,hh76,dnp76,wgu76} didn't have the direct
connection with the view of tunneling. Moreover, these methods, in which the
background geometry is considered fixed, didn't enforce the energy
conservation during the radiation process. Recently, Parikh and Wilczek
suggested \cite{pw00} a method based on energy conservation by calculating the
particle flux in Painlev\'{e} coordinates from the tunneling picture. Their
result can recover some important points and present some new properties as in
the following aspects: (1) It recovered the Hawking's original result in
leading order and gave the consistent temperature expression. (2) It gave a
direct relation between tunneling probability and the black hole entropy. (3)
The non-thermal spectrum shows there exists the information-carrying
correlation in the radiation, and the correlation can eliminate the entropy
growth completely and thus makes the process unitary \cite{zcz092}. The method
had been discussed generally in different situations
\cite{ecv01,zz05,rzg05,jwc06,jkf09,cnvzz07,km07} and its self-consistency has
been checked even by using the black hole thermodynamic law
\cite{sk08,tp08,zcz08,zcz091}.

In the calculation of tunneling, the choice of coordinates is very important
and the general discussion about it had been made in Ref. \cite{chnvz09}. It
is noted that a simple new coordinate system for de Sitter space was found in
Ref. \cite{kh02} and the tunneling through de Sitter horizon is generalized to
many other situations \cite{ajmm02,ss03}. It is demonstrated
\cite{kh02,ajmm02} that the coordinates can be used to calculate the tunneling
rate like that by Parikh and Wilczek, and the tunneling rate is non-thermal
when self-gravitation is taken into account. However, it is noted that the
entropy of empty de Sitter space is larger than that of Schwarzschild-de
Sitter space generally \cite{ss03,rb02}, which seems to show the tunneling of
de Sitter space was impossible since such tunneling leads to the transition of
empty de Sitter space to Schwarzschild-de Sitter space which challenges the
second law of thermodynamics. In order to clarify the problem, we
reinvestigate the tunneling radiation of Schwarzschild-de Sitter space and
discuss the entropy change in the process in the paper. We show the existence
of the correlation in the non-thermal spectrum of Schwarzschild-de Sitter
radiation, which explains the increase of entropy in the transition process of
Schwarzschild-de Sitter space to de Sitter space. Thus the total entropy,
contained the information about the matter in the space, the space itself and
their correlations, is conserved, which shows the tunneling radiation of de
Sitter space is possible and does not conflicted with the second law of thermodynamics.

\section{Tunneling through Schwarzschild-de Sitter Horizon Revisited}

To describe across-horizon phenomena, it is better to choose Painlev\'{e}-type
coordinates coordinates as obtained in Ref. \cite{kh02},%

\begin{equation}
ds^{2}=-(1-\frac{r^{2}}{l^{2}})dt^{2}-2\frac{r}{l}dtdr+dr^{2}+r^{2}d\Omega^{2}
\label{edsc}%
\end{equation}
where $l$ is the de Sitter spatial parameter with units of length called as
the de Sitter radius, which is the location of the horizon obtained by
$g_{tt}=0$.

However, it is noted that the tunneling rate obtained according to the metric
(\ref{edsc}) is thermal because such calculation does not include the
self-gravitation or back reaction \cite{kh02}. In order to incorporate the
self-gravitation effect, the Schwarzschild-de Sitter coordinates are
introduced to calculate the tunneling rate in Ref. \cite{kh02}. Historically,
the Schwarzschild-de Sitter coordinates played an important role in the work
of Gibbons and Hawking \cite{gh77} to determine the entropy of de Sitter
space, and it is also significant for our discussion here. Firstly, we
consider the three dimensional Schwarzschild-de Sitter coordinates
\cite{dj84,ssv01},
\begin{equation}
ds^{2}=-\left(  1-8GE-\frac{r^{2}}{l^{2}}\right)  dt_{s}^{2}+\frac{dr^{2}%
}{1-8GE-\frac{r^{2}}{l^{2}}}+r^{2}d\phi^{2}. \label{sdm}%
\end{equation}
In three dimensions there is only one horizon, at $r_{H}=l\sqrt{1-8GE}$, and
as $E$ goes to zero this reduces to the usual horizon in empty de Sitter
space. Here, the energy $E$ should be comprehended as the total energy existed
in the Schwarzschild-de Sitter space \cite{dj84,ssv01,bbm01}.

With the time Painlev\'{e} transformation, the corresponding Painlev\'{e}
coordinate is gotten as%
\begin{equation}
ds^{2}=-\left(  1-8GE-\frac{r^{2}}{l^{2}}\right)  dt^{2}-2\sqrt{8GE+\frac
{r^{2}}{l^{2}}}dtdr+dr^{2}+r^{2}d\phi^{2}%
\end{equation}
Considering the particle with energy $\omega$ tunneling across the horizon and
the self-gravitating effect, one can obtain the imaginary part of the action
\begin{equation}
\operatorname{Im}I=\operatorname{Im}\int_{E}^{E-\omega}\int_{r_{i}}^{r_{f}%
}\frac{drdH}{\overset{.}{r}}=-\operatorname{Im}\int_{0}^{\omega}\int_{r_{i}%
}^{r_{f}}\frac{drd\omega}{\sqrt{8G\left(  E-\omega^{^{\prime}}\right)
+\frac{r^{2}}{l^{2}}}-1}%
\end{equation}
where the radial outgoing geodesic with self-gravitation included is
$\overset{.}{r}=\sqrt{8G\left(  E-\omega^{^{\prime}}\right)  +\frac{r^{2}%
}{l^{2}}}-1$, which is obtained by setting $ds^{2}=d\phi^{2}=0$. Here
$r_{i}=l\sqrt{1-8GE}$ is the original radius of the horizon before
pair-creation, while $r_{f}=l\sqrt{1-8G\left(  E-\omega\right)  }$ is the new
radius of the horizon. Now the integral can be done by deforming the contour
according to the Feynman prescription,%

\begin{equation}
\operatorname{Im}I=\frac{\pi l}{4G}\left(  \sqrt{1-8GE}-\sqrt{1-8G\left(
E-\omega\right)  }\right)
\end{equation}
and then we can gain the tunneling rate%

\begin{equation}
\Gamma\sim\exp\left[  \frac{\pi l}{2G\hbar}\left(  \sqrt{1-8G\left(
E-\omega\right)  }-\sqrt{1-8GE}\right)  \right]  \label{dsc}%
\end{equation}
The tunneling rate we gained above is different from that in Ref. \cite{kh02},
$\Gamma\sim\exp\left[  \frac{\pi l}{2G\hbar}\left(  \sqrt{1-8GE}-1\right)
\right]  $ because they are obtained from different physical situations. This
point can be seen in the discussion of the next section.

\section{Thermodynamics and entropy}

Considering the tunneling particle's energy $\omega$ is small, we expand
$\Gamma$ given by Eq. (\ref{dsc}) in power of $\omega$ as%

\begin{equation}
\Gamma\sim\exp\left(  \frac{2\pi l\omega}{\hbar\sqrt{1-8GE}}+O\left(
\omega\right)  \right)
\end{equation}
where $\sqrt{1-8G\left(  E-\omega\right)  }=\sqrt{1-8GE}\sqrt{1+\frac
{8G\omega}{1-8GE}}\simeq\sqrt{1-8GE}\left(  1+\frac{4G\omega}{1-8GE}\right)
$. Compare with the Boltzmann factor, we find the temperature is given as%

\begin{equation}
T_{SdS}=\frac{\hbar\sqrt{1-8GE}}{2\pi l} \label{sdst}%
\end{equation}
which is consistent with that obtained in Ref. \cite{ssv01}. On the other
hand, if we calculated the surface gravity at the horizon, the temperature
obtained by $\frac{\kappa}{2\pi}$ is identical to that in Eq. (\ref{sdst}).
After the emission of energy $\omega$, the temperature becomes $T_{SdS}%
^{^{\prime}}=\frac{\hbar\sqrt{1-8G\left(  E-\omega\right)  }}{2\pi l}$, which
is higher than before. Since the temperature increases due to the tunneling,
so it will approach to $T_{dS}=\frac{\hbar}{2\pi l}$ and the Schwarzschild-de
Sitter space will decay to the empty de Sitter space quickly. In what follows,
we will give a physical picture of our calculation.

An observer locates at the origin while the tunneling occurs at the
Schwarzschild-de Sitter horizon. When the tunneling is proceeding, the horizon
becomes large gradually. It looks as if it were contrasted with the situation
for Schwarzschild black hole \cite{pw00} whose horizon contracts due to
particle tunneling. But if we consider it carefully, we will find that they
are consistent essentially because the horizon for both situations goes away
from the observer. In other words, the tunneling has the opposite direction
with the motion of horizon, which is consistent with energy conservation.
Finally the horizon becomes $r_{H}=l$ which is the radius of empty de Sitter
space and hence the Schwarzschild-de Sitter space decays into the empty de
Sitter space. Then is the de Sitter space the terminal of the tunneling? This
problem is difficult to answer because the empty de Sitter space has
thermodynamics with the temperature $T_{dS}=\frac{\hbar}{2\pi l}$. However,
more recently de Sitter space has entered cosmology as the likely candidate
for the final fate of the universe \cite{sl05}. Even though the de Sitter
space is regarded as the spacetime geometry of the early universe in rapid
inflation, its change to other stage has to be dependent on the new process
like reheating. From this opinion, we can also say the empty de Sitter space
is the terminal of the tunneling process. On the other hand, our picture is
analogous to the expanding universe. The qualitative analysis is made along
the line of the tunneling. Because the temperature is increasing gradually,
the tunneling rate will becomes quicker and quicker. In other words, the
horizon will become larger and larger and the velocity of the change will be
quicker and quicker. This is similar to accelerated expanding universe.

Using the thermodynamic relation,%
\begin{equation}
\frac{1}{T}=\frac{dS}{dM}=-\frac{dS}{dE}%
\end{equation}
where $dM=-dE$ in de Sitter space \cite{dse}. Thus one can find the entropy is
equal to%
\begin{equation}
S_{SdS}=\frac{\pi l}{2G\hbar}\sqrt{1-8GE}. \label{sds}%
\end{equation}
Thus we get the important result presented in Ref. \cite{pw00},%
\begin{equation}
\Gamma\sim\exp\left(  \Delta S\right)
\end{equation}
which shows that the tunneling rate is closely related to the change of the
entropy of Schwarzschild-de Sitter space. Note that besides the spatial
entropy, there may also be other entropy due to the energy $E$, $S(E)$, which
satisfies the relation $S_{SdS}+S(E)\leq S_{dS}=\frac{\pi l}{2G\hbar}$
\cite{rb02}. Therefore in the process of transition from Schwarzschild-de
Sitter space to de Sitter space the entropy is increased, but in Ref.
\cite{kh02,ajmm02} the tunneling in de Sitter space implies the entropy is
decreased since they discuss the inverse process. Although the two processes
are possible \cite{ajmm02}, they seem inconsistent with the second law of
thermodynamics which allows only the evolution toward one direction. In order
to solve this problem, we have to prove the entropy is conserved. In what
follows, we pay our attention to the non-thermal property of the tunneling
rate (\ref{dsc}) and show that there exist correlation in the tunneling
process along the line given in Ref. \cite{zcz092,zcz093}.

For two emissions with energies $\omega_{1}$and $\omega_{2}$, we find that%

\begin{equation}
\ln\Gamma(\omega_{1},\omega_{2})-\ln\left[  (\Gamma(\omega_{1})\ \Gamma
(\omega_{2})\right]  \neq0.
\end{equation}
Thus the adoption of a Schwarzschild-de Sitter tunneling does not change the
statement that a non-thermal spectrum affirms the existence of correlation, as
is illustrated in Ref. \cite{zcz092} for a Schwarzschild black hole. For
sequential tunneling of two particles with energies $\omega_{1}$ $\omega_{2}$,
we find the entropy form,%

\begin{align}
S(\omega_{1})  &  =-\ln\Gamma(\omega_{1})=-\frac{\pi l}{2G\hbar}\left(
\sqrt{1-8G\left(  E-\omega_{1}\right)  }-\sqrt{1-8GE}\right) \\
S(\omega_{2}|\omega_{1})  &  =-\ln\Gamma(\omega_{2}|\omega_{1})=-\frac{\pi
l}{2G\hbar}\left(  \sqrt{1-8G\left(  E-\omega_{1}-\omega_{2}\right)  }%
-\sqrt{1-8G\left(  E-\omega_{1}\right)  }\right)
\end{align}
and they also satisfy the definition of conditional entropy $S(\omega
_{1},\omega_{2})=-\ln\Gamma(\omega_{1},\omega_{2})=S(\omega_{1})+S(\omega
_{2}|\omega_{1})$. A detailed calculation confirms that the amount of
correlation is exactly equal to the mutual information described by
$S(A:B)=S\left(  A\right)  +S\left(  B\right)  -S\left(  A,B\right)  =S\left(
A\right)  -S\left(  A|B\right)  $ and this also shows that the correlation can
carry away the information. If we count the total entropy carried away by the
outgoing particles, we get
\begin{equation}
S(\omega_{1},\omega_{2},...,\omega_{n})=\sum\limits_{i=1}^{n}S(\omega
_{i}|\omega_{1},\omega_{2},\cdots,\omega_{i-1}).
\end{equation}
After calculation, we found that $S(\omega_{1},\omega_{2},...,\omega
_{n})=S_{dS}-S_{SdS}$. It seems that the entropy could increase in the process
since $S(E)\leq S(\omega_{1},\omega_{2},...,\omega_{n})$. However, we can
understand the situation by considering the radiation from a Schwarzschild
black hole. As described in Ref. \cite{zczy10}, we find the entropy of the
Schwarzschild black hole includes three parts: respectively associated with
the information for the precollapsed configurations, self-collapsed
configurations, and inter-collapsed configurations. Similarly, we can say in
the tunneling process of Schwarzschild-de Sitter space, the entropy is also
constituted of three parts: the spatial entropy (that is the entropy $S_{dS}$
or $S_{SdS}$), the matter entropy (that is the entropy $S(E)$) and the
correlation entropy. So it is the correlation that balance the entropy
inequality before and after tunneling and make the process unitary.

When the tunneling finishes, the correlations enlighten a way to explain the
entropy in empty de Sitter space. As discussed above, the energy is conserved,
so the energy $E$ of Schwarszchild-de Sitter space will change into the vacuum
energy of larger empty de Sitter space. Thus the entropy which includes the
Schwarzschild-de Sitter space entropy, the matter entropy and the correlation
entropy changes into the entropy of empty de Sitter space. Further, in empty
de Sitter space we can give some speculative suggestion that the ground states
or the vacuum states must be degenerate, which is similar to the explanation
of extreme charged black hole entropy \cite{sl05}. So the correlations could
be hidden among these degenerated states and the hidden correlation (the
inaccessible information in the empty de Sitter space for the observer located
in the origin) is the reason of spatial entropy increase, which shows that the
tunneling process is still unitary in principle. On the other side, we can say
that the entropy of empty de Sitter space is derived from the degeneracy of
ground states due to Boltzmann's microscopical explanation of entropy (the
microscopical explanation could be made along the line made for the
Schwarszchild black hole as in Ref. \cite{zcz092}). Therefore, we demonstrate
qualitatively why Schwarszchild-de Sitter coordinates are used to determine
the entropy of empty de Sitter space by Gibbons and Hawking \cite{gh77}.

\section{The High-dimensional Situation}

In the section we will discuss the situation of the tunneling radiation in
higher-dimensional Schwarzchild-de Sitter space-time. The line element in n+2
dimensional Schwarzschild-de Sitter spacetime can be expressed as
\cite{ajmm02}%

\begin{equation}
ds^{2}=-(1-\frac{r^{2}}{l^{2}}-\frac{M\epsilon_{n}}{r^{n-1}})dt^{2}%
+(1-\frac{r^{2}}{l^{2}}-\frac{M\epsilon_{n}}{r^{n-1}})^{-1}dr^{2}+r^{2}%
d\Omega_{n}^{2}%
\end{equation}
where $\epsilon_{n}=16\pi G_{n+2}/n\nu_{n}$ and $M$ is the conserved mass. It
is straightforward to generalize the prior calculation to the high-dimensional
situation and one can obtain that the tunneling rate is also non-thermal and
is related to the change of the entropy of Schwarzschild-de Sitter space
\cite{ajmm02}. Especially, there exists the black hole horizon in high
dimension besides the de Sitter horizon, but their tunneling or Hawking
radiation does not affect each other. For the black hole, the particle will
tunnel out of it, so its horizon shrinks; for de Sitter space, the particle
will tunnel into it, which is the same process as discussed in three
dimensional situation, so its horizon expands. Since the observer is
sandwiched between the two horizons, the two horizons move away from the
observer and the energy conservation is held. On the other hand, the
temperature of the two horizons is not equal to each other. The temperature at
the black hole horizon is larger, so there is net heat flow from the black
hole horizon to the de Sitter horizon, which will stimulate the black hole
radiation. Thus the black hole will disappear very soon. Moreover, in the
expanding process of de Sitter space caused by the tunneling through de Sitter
horizon, all matters such as black hole radiation and so on existed in the
space will change into the vacuum energy of the new space. Finally, the
high-dimensional Schwarzschild-de Sitter space will evolve into empty de
Sitter space.\textbf{ }

Is the entropy conserved in such process? The answer is positive although the
process is more intricate than that in three dimensional situation. Since
there are two horizon in the high-dimensional Schwarzschild-de Sitter space,
the equivalent Hawking radiation could be described as that in the Ref.
\cite{ss03} where the entropy relation is given as%

\begin{equation}
S_{h}+S_{c}+2\sqrt{S_{c}S_{h}}=S_{SdS}<S_{dS}%
\end{equation}
where $S_{h}$, $S_{c}$ are the entropies of event and cosmological horizons,
respectively and the term $2\sqrt{S_{c}S_{h}}$ may be the correlation between
the two horizons, and they together constitute the entropy of high-dimensional
Schwarzschild-de Sitter space. However, when the radiation happens, there is
another neglected kind of entropy related to the correlation between the
horizons and the radiation, which can be obtained according to the same method
in the last section. It is such correlations that balance the inequality of
the entropy of Schwarzschild-de Sitter space and the de Sitter space, which
implies even the tunneling process in the high-dimensional Schwarzschild-de
Sitter space is also unitary. Therefore, the entropy of the empty de Sitter
space could be described as deriving from the entropy of Schwarzschild-de
Sitter space and the correlation hidden in the vacuum.

\section{\textbf{Conclusion}}

In conclusion, we have presented the transition of Schwarszchild-de Sitter
space to de Sitter space through tunneling, which is consistent with the
accelerated expanding universe and shows the stability of de Sitter space
indirectly. We also show that there exists the correlation in the non-thermal
spectrum and the tunneling process is entropy-conserved. Thus the entropy of
empty de Sitter space is explained logically from the correlations and the
initial entropy taken by matter and the space itself, which shows physically
the feasibility of the introduction of Schwarszchild-de Sitter coordinates to
determine the entropy of empty de Sitter space by Gibbons and Hawking.

\section{Acknowledgement}

This work is supported by National Natural Science Foundation of China under
Grant No. 11074283 and 11104324.

\end{document}